\documentclass[preprint,superscriptaddress,amsmath,amssymb,prd,aps,showpacs,floatfix,nofootinbib]{revtex4-1}
\usepackage{graphicx}% Include figure filesa
\usepackage{dcolumn}% Aligntable columns on decimal point
\usepackage{bm}
\usepackage{mathrsfs}% bold math
\usepackage{hyperref}% add hypertext capabilities
\usepackage{amsmath}
\usepackage{amssymb}
\usepackage{bm}
\usepackage{color}
\usepackage{float}
\usepackage{dcolumn}
\usepackage{multirow}
\usepackage{changepage}
\usepackage{enumerate}
\usepackage[normalem]{ulem}
\usepackage[dvipsnames]{xcolor}
\usepackage{braket}
\usepackage{nicefrac}
\usepackage{multirow}
\usepackage{tabularx}
\usepackage{subfigure}
\usepackage{orcidlink}

\hypersetup{pdftex,colorlinks=true,linkcolor=blue,citecolor=red,menucolor=black,urlcolor=blue,filecolor=blue}

\newcommand{\mev}{\textrm{ MeV}}
\newcommand{\gev}{\textrm{ GeV}}

\begin{document}
%\title{$\boldsymbol{\Xi_{bb}}$ and $\boldsymbol{\Omega_{bbb}}$ molecular states}
\title{\boldmath $J/\psi$ decays into $\omega (\phi) f_1(1285)$ and $\omega (\phi) ``f_1(1420)"$ }
\date{\today}

\author{Jia-Xin Lin}
\affiliation{Department of Physics, Guangxi Normal University, Guilin 541004, China}
\affiliation{School of Physics, Southeast University, Nanjing 210094, China}

\author{Jia-Ting Li}
\affiliation{Department of Physics, Guangxi Normal University, Guilin 541004, China}

\author{Wei-Hong Liang}
%\author{Wei-Hong Liang \orcidlink{0000-0001-5847-2498}}
\email{liangwh@gxnu.edu.cn}
\affiliation{Department of Physics, Guangxi Normal University, Guilin 541004, China}
\affiliation{Guangxi Key Laboratory of Nuclear Physics and Technology, Guangxi Normal University, Guilin 541004, China}

\author{Hua-Xing Chen}
%\author{Hua-Xing Chen\orcidlink{0000-0002-5141-6888}}
\email{hxchen@seu.edu.cn}
\affiliation{School of Physics, Southeast University, Nanjing 210094, China}

\author{Eulogio Oset}
%\author{Eulogio Oset \orcidlink{0000-0002-4462-7919}}
\email{oset@ific.uv.es}
\affiliation{Department of Physics, Guangxi Normal University, Guilin 541004, China}
\affiliation{Departamento de F\'{\i}sica Te\'orica and IFIC, Centro Mixto Universidad de
Valencia-CSIC Institutos de Investigaci\'on de Paterna, Aptdo.22085,
46071 Valencia, Spain}

%             %  but any date may be explicitly specified

\begin{abstract}
We perform a theoretical study of the $J/\psi \to \omega (\phi) K^* \bar{K} + c.c. \to \omega (\phi) K^0 \pi^+ K^-$ reactions with the assumption that the $f_1(1285)$ is dynamically generated from a single channel $K^* \bar{K} + c.c$ interaction in the chiral unitary approach. Two peaks in the $K^0 \pi^+ K^-$ invariant mass distribution are observed, one clear peak locates at the $f_1(1285)$ nominal mass, the other peak locates at around $1420 \mev$ with about $70 \mev$ width. We conclude that the former peak is associated with the $f_1(1285)$ and the latter peak is not a genuine resonance but a manifestation of the kinematic effect in the higher energy region caused by the $K^* \bar{K} + c.c.$ decay mode of the $f_1(1285)$.

\end{abstract}

%\pacs{11.30.Er, 12.39.-x, 13.25.Hw}% PACS, the Physics and Astronomy
\maketitle

%=====================
%=====================
%=====================
\section{Introduction}
\label{sec:intro}
%=====================
%=====================
%=====================

The study of the internal structure and properties of hadrons, and the mechanisms of strong interaction between hadrons is an important subject in the field of hadron physics.
The axial vector mesons provide a powerful source of information about hadron dynamics.
The $f_1(1285)$ is an axial vector meson with quantum numbers $I^G(J^{PC})=0^+(1^{++})$, which was discovered in 1960s \cite{Miller:1965zza, dAndlau:1965cww, Dahl:1967pg, DAndlau:1968vje}.
The mass and the total decay width of the $f_1(1285)$ are $(1281.9 \pm 0.5) \mev$ and $(22.7 \pm 1.1) \mev$, respectively \cite{Workman:2022ynf}.
The $f_1(1285)$ resonance is described as a $q\bar{q}$ state within the quark model \cite{Gavillet:1982tv, Godfrey:1998pd, Li:2000dy, Klempt:2007cp, Vijande:2004he, Chen:2015iqa}.
It is also claimed to be a $K^*\bar{K}$ molecular state in Refs.~\cite{Lutz:2003fm, Roca:2005nm}, dynamically generated from the interaction of a single channel $K^* \bar{K} + c.c.$ in the chiral unitary approach.
Since the $f_1(1285)$ is about $100 \mev$ below the $K^*\bar{K}$ threshold, the $K^*\bar{K}$ decay mode is difficult to observe.
Indeed, the dominant decay modes for the $f_1(1285)$ are $4\pi$, $\eta \pi \pi$, and $K \bar{K} \pi$, with their  corresponding branching fraction being $32.7\%$, $35.0\%$, and $9.0\%$, respectively \cite{Workman:2022ynf}.

It is proper to stress the difference of the two mentioned pictures for the $f_1(1285)$. In the quark models one sticks to the conventional structure of the mesons as $q \bar q$ objects (although one can also address more complicated structures as tetraquarks). What we call dynamically generated states, are states that emerge from the interaction of more elementary particles, one pseudoscalar and one vector meson in the present case, forming bound states or resonances from the interaction (potential from the Schroedinger equation point of view). One refers equally to the dynamically generated states as molecular states, and the emerging states would be some kind of bound states of these interacting particles, much as the deuteron would be a bound state of a proton and the neutron. There is an immediate difference between these molecular states and the $q \bar q$ states, since in the molecular picture one already has four quarks (two quarks and two antiquarks). Quark models with four quarks can have overlap with a molecular structure and there are works in this direction \cite{david,alessandro}. Even then, one differentiates between compact quark states, coming from the direct quark interaction, and the molecular states that emerge from a residual interaction between the two quark clusters of the two mesons.

The work of Ref.~\cite{Roca:2005nm} was extended to include higher order terms in the Lagrangian in Ref.~\cite{Zhou:2014ila}, but the results showed that the effect of the higher order kernel could be neglected.
This work was also applied in finite volume to study the $s$\,-wave $K K^*$ interaction in the $f_1(1285)$ channel \cite{Geng:2015yta}.
Besides, many theoretical works have been done within the molecular picture of the $f_1(1285)$ \cite{Roca:2005nm}.
The $f_1(1285) \to K \bar{K} \pi$ decay \cite{Aceti:2015pma} and the $f_1(1285) \to \eta \pi^0 \pi^0$ decay \cite{Aceti:2015zva} were studied, and their branching ratios are compatible with the experimental results \cite{Workman:2022ynf, WA102:1997gkz, WA102:1998zhh}.
The theoretical results of Ref.~\cite{Aceti:2015zva} were confirmed later in the BES\uppercase\expandafter{\romannumeral3} experiment \cite{BESIII:2015you}.
The $K^- p \to f_1(1285) \Lambda$ decay has been investigated using the effective Lagrangian approach in Ref.~\cite{Xie:2015wja}, which reproduced the experimental measurement in Ref.~\cite{Amsterdam-CERN-Nijmegen-Oxford:1978mws}.
The radiative decay $f_1(1285) \to \gamma \rho^0 (\omega, \phi)$ was studied in Ref.~\cite{Xie:2019iwz}, and the decay width $\Gamma_{f_1 \to \gamma \rho^0}$ obtained is in agreement with the experimental result \cite{CLAS:2016zjy}.
Recently, the ratio $\Gamma_{\bar{B}^0 \to J/\psi \bar{K}^{*0} K^0}/\Gamma_{\bar{B}^0 \to J/\psi f_1(1285)}$ related to the $\bar{B}^0 \to J/\psi \bar{K}^{*0} K^0$ and $\bar{B}^0 \to J/\psi f_1(1285)$ decays was predicted in Ref.~\cite{He:2021exv}.

Compared with the $f_1(1285)$ resonance, the nature and the structure of the $f_1(1420)$ resonance are more controverial. It has mostly been found in high energy reactions, such as $pp$ \cite{WA76:1989zfi, E690:1997tev}, $\pi^- p$ \cite{Prokoshkin:1997ze, CERN-CollegedeFrance-Madrid-Stockholm:1980umk, E852:2001ote}, and $e^+ e^-$ decays \cite{MARK-III:1986unh, DELPHI:2003bnm}.
The $f_1(1420) \to K^* \bar{K}$ was first reported in the $\pi^- p \to K_s^0 K^{\pm} \pi^{\pm} n$ reactions at $3.95 \gev/c$ \cite{CERN-CollegedeFrance-Madrid-Stockholm:1980umk}.
Then the $f_1(1420) \to K \bar{K} \pi$ was observed in the $J/\psi \to \gamma K_S^0 K^\pm \pi^\mp$ reactions \cite{MARK-III:1990wgk}.
The $f_1(1420)$ is catalogued as a standard resonance in the PDG with the quantum numbers $I^G(J^{PC})=0^+(1^{++})$.
The mass and the total decay width of the $f_1(1420)$ are $(1426.3 \pm 0.9) \mev$ and $(54.5 \pm 2.6) \mev$, respectively \cite{Workman:2022ynf}.
Currently, the $f_1(1420)$ is only seen in the $K \bar{K} \pi$, $a_0(980) \pi$, and $\phi \gamma$ decay modes.

Yet, the classification of the $f_1(1420)$ as a standard resonance has been challenged in some theoretical works.
In the work of Ref.~\cite{Debastiani:2016xgg}, the $f_1(1285) \to \pi a_0(980)$ and $f_1(1285) \to K^* \bar{K}$ were studied. A clear peak around $1285 \mev$ was observed in both decay modes, but a shoulder was found at about $1400 \mev$ for the former decay mode and a peak around $1420 \mev$ with a width about $60 \mev$ was found for the latter decay mode.
On the theoretical side, the results of Ref.~\cite{Debastiani:2016xgg} showed that the shoulder of the $a_0(980) \pi$ decay mode was caused by a triangle singularity.
Moreover, the structure of the $f_1(1420)$ was tested again in Ref.~\cite{Liang:2020jtw}, and a peak around $1420 \mev$ of the $K\bar{K}\pi$ invariant mass was also found, with this peak corresponding to the $f_1(1285) \to K^* \bar{K}$ decay with the $K^*$ placed on shell at higher invariant masses.

It is worth to elaborate further into this issue and the relationship of the $f_1(1420)$ to the triangle singularity. 
There was some excitement when the COMPASS collaboration announced the discovery of the $a_1(1420)$ resonance, but soon it was dismissed \cite{misha,fcaa1} since the signal, seen in the decay to $\pi f_0(980)$, emerged as a consequence of a triangle singularity \cite{landau} (see also the same conclusion reached by the COMPASS group \cite{COMPASS:2020yhb}), 
concretely from a triangle loop mechanism where another resonance, the $a_1(1260)$, decays to $K^* \bar K$, subsequently the $K^*$ decays to $\pi K$ and the $K \bar K$ merge to give $f_0(980)$. 
The singularity in the triangle diagram appears when all the particles in the loop are placed on shell simultaneously and they are collinear, such that the process can occur at the classical level \cite{norton} (see a pedagogical reformulation of the problem in Ref.~\cite{bayarguo}). 
The $f_1(1420)$ has a reported decay mode into $\pi a_0(980)$, analogous to the former one, which however is interpreted in Ref.~\cite{Debastiani:2016xgg} as the $f_1(1285)$ decaying to $K^* \bar K$, the $K^*$ decaying then to $\pi K$ and the $K \bar K$ merging to give the $a_0(980)$.  
Yet, the big signal of the $f_1(1420)$ is seen in the $K^* \bar K$ decay channel, which is interpreted in Ref.~\cite{Debastiani:2016xgg} as the consequence  of the opening of the $K^* \bar K$ channel in the $f_1(1285)$ decay. 
There are, hence, some analogies but also differences between the ``$a_1(1420)$" and the ``$f_1(1420)$".

The BES\uppercase\expandafter{\romannumeral3} Collaboration measured the $J/\psi \to \omega \eta \pi^+ \pi^-$ decay and found a clear peak around $1285 \mev$ with a width around $22 \mev$ in the $\eta \pi^+ \pi^-$ mass distribution, which was associated to the $f_1(1285)$ production \cite{BESIII:2011nqb}.
The product branching fraction $\mathcal{B}[J/\psi \to \omega f_1(1285)] \times \mathcal{B}[f_1(1285) \to a_0(980)^\pm \pi^\mp] \times \mathcal{B}[a_0(980)^\pm \to \eta \pi^\pm] = (1.25 \pm 0.10_{-2.0}^{+1.9}) \times 10^{-4}$ was also given in Ref.~\cite{BESIII:2011nqb}.
The observation of the $f_1(1285)$ signal in Ref.~\cite{BESIII:2011nqb} motivates us to study $J/\psi \to \omega (\phi) f_1(1285)$ and $J/\psi \to \omega (\phi) K^* \bar{K} + c.c. \to \omega (\phi) K^0 \pi^+ K^-$ decays, testing the nature of the axial vector meson $f_1(1285)$ and trying to explain the origin and nature of the $f_1(1420)$ resonance.
We perform the study of the $J/\psi \to \omega (\phi) K^* \bar{K} + c.c. \to \omega (\phi) K^0 \pi^+ K^-$ decays with the assumption that the $f_1(1285)$ is dynamically generated from a single channel $K^*\bar{K}+c.c.$ interaction in the chiral unitary approach.
We observe two peaks in the $K^0 \pi^+ K^-$ invariant mass distribution: one peak locating around $1285 \mev$ which is associated to the $f_1(1285)$ resonance, the other peak locating around $1420 \mev$.
The latter peak is a manifestation of the kinematic effect caused by the $K^*\bar{K}+c.c.$ decay mode of the $f_1(1285)$ when the $K^*$ is placed on shell at higher invariant masses, and, thus, should not be attributed to a genuine resonance.

This paper is organized as follows.
In Sec.~\ref{sec:form} we discuss formalism and the main ingredients of the model.
Following the formalism we show the results in Sec.~\ref{sec:result}, and finally, we give a brief conclusion in Sec.~\ref{sec:concl}.

%=====================
%=====================
%=====================
\section{Formalism}
\label{sec:form}
%=====================
%=====================
%=====================
\subsection{$f_1 (1285)$ as a dynamically generated state}

We want to study the role of the $f_1(1285)$ state in the $J/\psi \to \omega f_1(1285)$ decay, which is dynamically generated from a single channel $K^* \bar{K} + c.c.$ interaction.
In the chiral unitary approach, the scattering amplitude of $f_1(1285)$ resonance could be obtained by solving the Bethe-Salpeter equation \cite{Roca:2005nm}
%%%%%%%%%%%%%%%%%
\begin{equation}
	\label{eq:BS}
	T=[1 + V \hat{G}]^{-1} \; (-V) \;\vec{\epsilon} \cdot \vec{\epsilon}^{~\prime},
\end{equation}
%%%%%%%%%%%%%%%%%
where $V$ is the transition potential, $\vec{\epsilon}$ ($\vec{\epsilon}^{~\prime}$) is the polarization vector of initial (final) vector meson, the  $\hat{G}_l = G_l (1+ \frac{1}{3}\frac{q_l^2}{M_l^2})$ is a diagonal matrix with the $l$th element, $q_l$ is given by
%%%%%%%%%%%%%%%%%
\begin{eqnarray}
	q_l = \frac{1}{2\sqrt{s}} \;\sqrt{[s-(M_l+m_l)^2]\;[s-(M_l-m_l)^2]},
\end{eqnarray}
%%%%%%%%%%%%%%%%%
and $G_l$ is the meson-meson loop function
%%%%%%%%%%%%%%%%%
\begin{eqnarray}
	\label{eq:Gl}
	G_l (\sqrt{s}) = i \int \frac{\mathrm{d}^4q}{(2\pi)^4} \;\frac{1}{(P-q)^2 - M_l^2 + i\epsilon} \;\frac{1}{q^2 - m_l^2 + i\epsilon},
\end{eqnarray}
%%%%%%%%%%%%%%%%%
conveniently regularized.
In the above expression, $M_l$ and $m_l$ are the masses of vector and pseudoscalar mesons, respectively, $P$ is the total four-momentum of these two mesons, $P^2=(\sqrt{s})^2$, with $\sqrt{s}$ being the centre-of-mass energy of these two mesons in the loop.
Since the loop function of Eq.~\eqref{eq:Gl} is divergent, we usually use either the cut-off method \cite{Oller:1997ti} or the subtraction constant in the dimensional regularization method \cite{Oller:2000fj} to renormalize it.
Here we use the cut-off method and then Eq.~\eqref{eq:Gl} becomes
%%%%%%%%%%%%%%%%%
\begin{eqnarray}
	\label{eq:G-cutoff}
	G_l(\sqrt{s}) = \int_{0}^{q_{\rm max}} \frac{q^2 \mathrm{d} q}{(2\pi)^2} \;\frac{\omega_1+\omega_2}{\omega_1\omega_2 \;[s-(\omega_1+\omega_2)^2 + i\epsilon]},
\end{eqnarray}
%%%%%%%%%%%%%%%%%
where $\omega_1=\sqrt{M_l^2 + \mathbf{q}}$, $\omega_2=\sqrt{m_l^2 + \mathbf{q}}$, and $q_{\rm max}$ is the maximum value of the modulus of the three momentum $\mathbf{q}$ allowed in the integral of Eq.~\eqref{eq:Gl}.
In Ref.~\cite{Roca:2005nm} the $f_1(1285)$ resonance was reproduced with a cutoff $q_{\rm max}=1000\mev$, which gave a good description of the $f_1(1285)$ resonance.
Hence we take the same value $q_{\rm max}=1000\mev$.
We consider the $K^*$ meson with and without width in Eq.~\eqref{eq:Gl}, obtaining a pole of the $T$ matrix at $(1281+i14)\mev$ and $(1281+i0)\mev$, respectively.
Note that we substitute $M_{K^*}^2 \to M_{K^*}^2 - i M_{K^*} \Gamma_{K^*}$ in Eq.~\eqref{eq:G-cutoff} for the former case.

Then the coupling of the $f_1(1285)$ resonance to the $K^*\bar{K} + c.c.$ channel in $s$-wave is obtained from the residues of the $T_{f_1,f_1}$ amplitude at the pole
%%%%%%%%%%%%%%%%%
\begin{eqnarray}
	\label{eq:coupling}
	T_{f_1,f_1}=\frac{g_{f_1, K^*\bar{K}}^2}{z-z_R},
\end{eqnarray}
%%%%%%%%%%%%%%%%%
where $z$ is the complex energy and $z_R$ is the complex pole position.
With the poles $(1281+i14)\mev$ and $(1281+i0)\mev$, we get the couplings as
%%%%%%%%%%%%%%%%%
\begin{eqnarray}
	\label{eq:gf1}
	g_{f_1,K^* \bar{K}} &=& (7474-i364) \mev, \\
	g_{f_1,K^* \bar{K}}^\prime &=& (7469-i0) \mev,
\end{eqnarray}
%%%%%%%%%%%%%%%%%
where $g_{f_1,K^* \bar{K}}$ and $g_{f_1,K^* \bar{K}}^\prime$ correspond to the cases with and without consideration of the width of the $K^*$ meson, respectively.
We shall use these two values in our calculation.

\subsection{Decay of $J/\psi \to \omega f_1 (1285)$, $f_1(1285) \to K^* \bar K +c.c.$}

\begin{figure}[b]
	\includegraphics[width=0.45\linewidth]{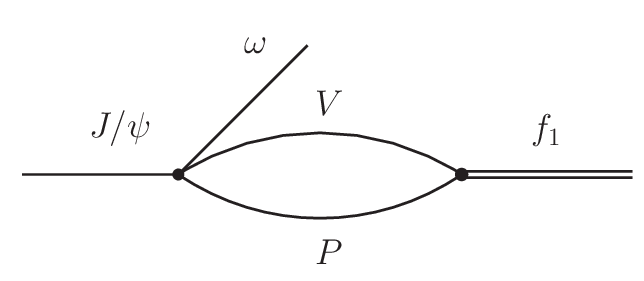}
	\vspace{-0.4cm}
	 \caption{The mechanism for the $J/\psi\to\omega f_1(1285)$ decay. }
	 \label{fig:omegaf1}
 \end{figure}

Within a molecular picture of the $f_1(1285)$, the decay of $J/\psi \to \omega f_1(1285)$ must proceed via the decay of $J/\psi \to \omega VP$, and the $f_1(1285)$ will be generated by the $VP$ final state interaction, where $V$ and $P$ stand for vector meson and pseudoscalar meson, respectively.
The mechanism for the $J/\psi\to\omega f_1(1285)$ decay is shown in Fig.~\ref{fig:omegaf1}.
%%%%%%%%%%%%%%%%%
%%%%%%%%%%%%%%%%%

We analyse the structure for the first vertex $J/\psi \omega VP$ in Fig.~\ref{fig:omegaf1}. 
Since the quark component of $J/\psi$ is $c \bar{c}$ without any light quarks, the $J/\psi$ is an SU(3) singlet. Then the remaining particles $\omega VP$ are constructed as an SU(3) singlet.
There are four structure forms for $\omega VP$, $\it{i.e.}$, $\braket{VVP}$, $\braket{V} \braket{VP}$, $\braket{VV} \braket{P}$ and $\braket{V}\braket{V}\braket{P}$, where the symbol $\langle~\rangle$ denotes the trace of a matrix. The vector and pseudoscalar SU(3) matrices are given by
%%%%%%%%%%%%%%%%%
\begin{equation}\label{eq:V}
	V=
	\left(
	\begin{array}{ccc}
	\frac{1}{\sqrt{2}}\rho^0+\frac{1}{\sqrt{2}}\omega&\rho^+&K^{*+}\\
	\rho^-&-\frac{1}{\sqrt{2}}\rho^0+\frac{1}{\sqrt{2}}\omega&K^{*0}\\
	K^{*-}&\bar{K}^{*0}&\phi\\
	\end{array}
	\right),
\end{equation}
%%%%%%%%%%%%%%%%%
\begin{equation}\label{eq:P}
P=
\left(
\begin{array}{ccc}
\frac{1}{\sqrt{2}} \pi^0+ \frac{1}{\sqrt{3}} \eta + \frac{1}{\sqrt{6}} \eta^\prime & \pi^+ & K^+\\
\pi^- & -\frac{1}{\sqrt{2}} \pi^0 + \frac{1}{\sqrt{3}} \eta + \frac{1}{\sqrt{6}}\eta^\prime & K^0\\
K^- & \bar{K}^0 & -\frac{1}{\sqrt{3}} \eta + \sqrt{\frac{2}{3}} \eta^\prime
\end{array}
\right),
\end{equation}
%%%%%%%%%%%%%%%%%
where the $\eta-\eta'$ mixing is considered in $P$ matrix \cite{Bramon:1992kr}.
It is shown in Ref.~\cite{Manohar} that structures involving more traces are suppressed by the large $N_c$ counting. According to that, the dominant structure is $\braket{VVP}$, which was also found empirically to dominate in Refs.~\cite{Liang:2016hmr,Debastiani:2016ayp}.
Hence, we propose the structure  $\braket{VVP}$ as the dominant component of the $J/\psi VVP$ vertex,
%%%%%%%%%%%%%%%%%
\begin{equation}\label{eq:vertex}
H=\braket{VVP}.
\end{equation}
%%%%%%%%%%%%%%%%%
More details about this structure can be seen in Refs.~\cite{Liang:2016hmr,Debastiani:2016ayp}.
Using the $V$ and $P$ matrices of Eqs.~\eqref{eq:V} and \eqref{eq:P}, Eq.~\eqref{eq:vertex} gives the structure
%%%%%%%%%%%%%%%%%
\begin{equation}\label{eq:H1}
H=\frac{\omega}{\sqrt{2}}K^{*+}K^-+\frac{\omega}{\sqrt{2}}K^{*0}\bar{K}^0+K^{*-}\frac{\omega}{\sqrt{2}}K^++\bar{K}^{*0}\frac{\omega}{\sqrt{2}}K^0.
\end{equation}

Then we need to analyse the spin-parity structure of the first vertex of Fig.~\ref{fig:omegaf1}. The $I^G (J^{PC})$ quantum numbers of the $J/\psi$, $\omega$, and $f_1(1285)$ are $0^-(1^{--})$, $0^-(1^{--})$, and $0^+(1^{++})$, respectively.
The transition involves three vectors, $J/\psi$ and two extra vectors. From Eq.~\eqref{eq:H1} we would have $J/\psi \to V_1 \,V_2 \,P$. With the quantum numbers seen before, the vertex can proceed in $s$-wave, and we need to make a scalar object with the combination
\begin{equation}
	\label{eq:VVP1}
	\vec{\epsilon}\,(J/\psi) \cdot [\vec{\epsilon}\,(V_1)\times\vec{\epsilon}\,(V_2)] = -\,\vec{\epsilon}\,(J/\psi) \cdot [\vec{\epsilon}\,(V_2)\times\vec{\epsilon}\,(V_1)].
\end{equation}
%%%%%%%%%%%%%%%%%
Taking into account Eq.~\eqref{eq:VVP1}, we can reorder Eq.~\eqref{eq:H1} as

\begin{equation}\label{eq:H}
H=\frac{\omega}{\sqrt{2}}\left(K^{*+}K^-+K^{*0}\bar{K}^0-K^{*-}K^+-\bar{K}^{*0}K^0\right).
\end{equation}
%%%%%%%%%%%%%%%%%

Taking the same convention as Ref.~\cite{Roca:2005nm}, $\it{i.e.}$, $\ket{K^-}=-\ket{1/2,-1/2}$, $\ket{K^{*-}}=-\ket{1/2,-1/2}$, $C(K^{*+})=-K^{*-}$, we find that the kaon structure of Eq.~\eqref{eq:H} corresponds to the $f_1(1285)$ resonance of isospin $I=0$ and $C$-parity $C=+$,
%%%%%%%%%%%%%%%%%
\begin{equation}\label{eq:f1}
	\ket{f_1(1285)}=-\frac{1}{2}\ket{K^{*+}K^-+K^{*0}\bar{K}^0-K^{*-}K^+-\bar{K}^{*0}K^0}.
\end{equation}

\subsection{The $J/\psi \to \omega K^0 \pi^+ K^-$ decay}

In the experiment, the $f_1(1285)$ state was observed in the $K\bar K\pi$ invariant mass distribution. To have $K\bar K \pi$ in the final state, we make one more step to consider the $K^*$ or $\bar K^*$ decay,
%%%%%%%%%%%%%%%%%
\begin{equation}
\label{eq:decay1}
J\psi\to\omega K^{*+}K^-, ~~K^{*+}\to K^0\pi^+,
\end{equation}
%%%%%%%%%%%%%%%%%
or
%%%%%%%%%%%%%%%%%
\begin{equation}
\label{eq:decay2}
J/\psi\to\omega \bar{K}^{*0}K^0,~~\bar{K}^{*0}\to K^-\pi^+.
\end{equation}
%%%%%%%%%%%%%%%%%

%%%%%%%%%%%%%%%%%
\begin{figure}[b]
\centering
\subfigure[Mechanisms for $J/\psi \to \omega K^{*+} K \to \omega K^0 \pi^+ K^-$ decay.]{
\label{fig-2}
\includegraphics[width=0.8\linewidth]{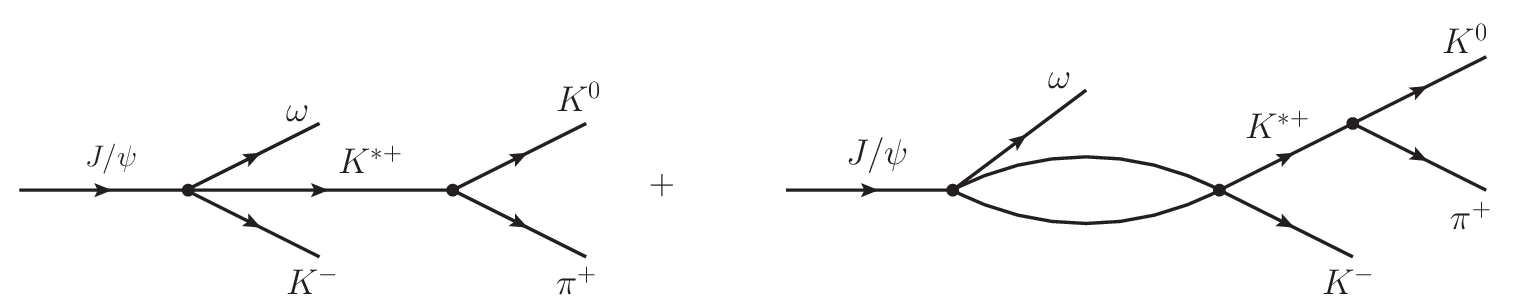}}
\subfigure[Mechanisms for $J/\psi \to \omega\bar{K}^{*0} K^0 \to \omega K^0 \pi^+ K^-$ decay.]{
\label{fig-3}
\includegraphics[width=0.8\linewidth]{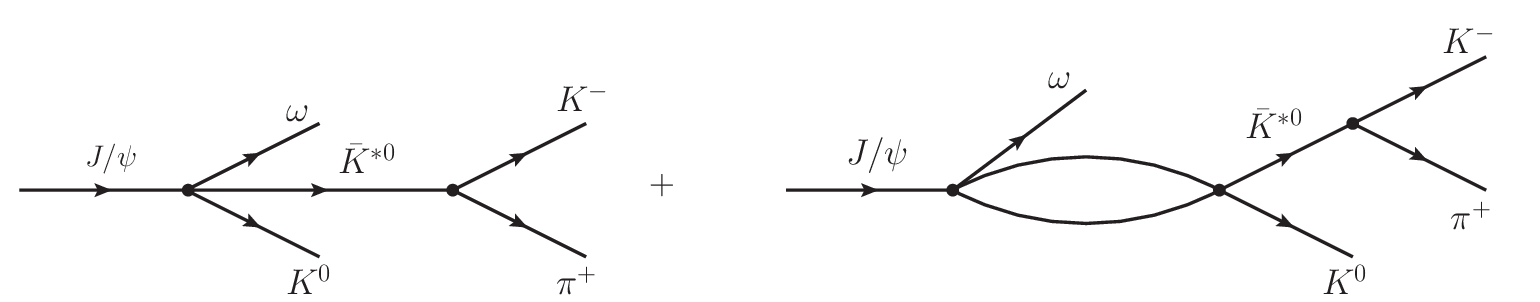}}
\caption{Mechanisms for $J/\psi \to \omega K^*\bar{K}+c.c. \to \omega K^0\pi^+K^-$ decays containing tree level diagrams (left) and rescattering diagrams of $VP$ $(K^*\bar{K}, \bar{K}^* K)$ components to produce the $f_1(1285)$ resonance (right).}
\label{fig:feynman}
\end{figure}
%%%%%%%%%%%%%%%%%

For the $J/\psi \to \omega K^*\bar{K} + c.c. \to \omega K^0\pi^+K^-$ decay, the mechanisms are shown in Fig.~\ref{fig:feynman}, which contain two decay processes: 1) tree level decay; 2) rescattering of $VP$ ($K^* \bar{K}, \bar{K}^* K$) producing the $f_1(1285)$ resonance that decays into $K^* \bar{K}$ or $\bar{K}^* K$ and later into $K^0 \pi^+ K^-$.
Then we need to consider the $K^{*+} K^0 \pi^+$ and $\bar{K}^{*0} K^- \pi^+$ vertices in Fig.~\ref{fig:feynman}.

The decay vertex in a $K^*$ decay to $K\pi$ is given by the standard Lagrangian
\begin{equation}
	{\mathcal{L}}= -ig \braket{\left[P, \partial_\mu  P\right] V^\mu};~~~ g=\frac{M_V}{2f},~M_V=800\,\mev,~ f=93\, \mev.
\end{equation}
Neglecting the $\epsilon^0$ component of the polarization vector of the slow moving $K^*$, as justified in Ref.~\cite{Sakai}, we find the couplings
\begin{equation}\label{eq:vectex1}
   g\, \vec \epsilon\, (K^{*+}) \cdot (\vec p_{\pi^+} -\vec p_{K^0}),~~{\rm for} ~ K^{*+} \to \pi^+ K^0,
\end{equation}
\begin{equation}\label{eq:vectex2}
	-g\, \vec \epsilon\, (\bar K^{*0}) \cdot (\vec p_{\pi^+} -\vec p_{K^-}),~~{\rm for} ~ \bar K^{*0} \to \pi^+ K^-,
 \end{equation}
with $\vec{p}_{\pi^+}$, $\vec{p}_{K^0}$, and $\vec{p}_{K^-}$ the momentum of $\pi^+$, $K^0$, and $K^-$, respectively.

Using the vertices of Eqs.~\eqref{eq:VVP1}, \eqref{eq:vectex1} and \eqref{eq:vectex2}, the amplitudes corresponding to the mechanisms of Fig.~\ref{fig-2} and Fig.~\ref{fig-3} are given by
%%%%%%%%%%%%%%%%%
\begin{eqnarray}
	\label{eq:amp1}
	t_a &=& \mathcal{C}\, \vec{\epsilon}\,(J / \psi) \cdot \left[\vec{\epsilon}\,(\omega) \times \vec{\epsilon}\,(K^{*+})\right] \vec{\epsilon}\,(K^{*+}) \cdot (\vec{p}_{\pi^{+}}-\vec{p}_{K^0}) \cdot D_{K^{*+}}\left(M_{\rm inv}(K^0 \pi^+)\right) \nonumber \\[1.5mm]
	&& + \mathcal{C}\, \vec{\epsilon}\,(J / \psi) \cdot\left[\vec{\epsilon}\,(\omega) \times \vec{\epsilon}\,(K^{*+})\right] \vec{\epsilon}\,(K^{*+}) \cdot (\vec{p}_{\pi^{+}}-\vec{p}_{K^0}) \cdot G_{K^*\bar{K}}\left( M_{\rm inv} (K^0 \pi^+ K^-)\right) \nonumber\\[1.5mm]
	&&\times T_{K^*\bar{K},K^*\bar{K}}^{(f_1)} \left(M_{\rm inv} (K^0 \pi^+ K^-)\right) \cdot D_{K^{*+}}\left(M_{\rm inv}(K^0 \pi^+)\right), \\[2mm]
	%-----------------------------------------------------------
	\label{eq:amp2}
	t_b &=& - \mathcal{C}\, \vec{\epsilon}\,(J / \psi) \cdot \left[\vec{\epsilon}\,(\omega) \times \vec{\epsilon}\,(\bar{K}^{*0})\right] \left[-\vec{\epsilon}\,(\bar{K}^{*0}) \cdot (\vec{p}_{\pi^{+}}-\vec{p}_{K^-})\right] \cdot D_{\bar{K}^{*0}}\left(M_{\rm inv}(K^- \pi^+)\right) \nonumber \\[1.5mm]
	&& + \mathcal{C}\, \vec{\epsilon}\,(J / \psi) \cdot \left[\vec{\epsilon}\,(\omega) \times \vec{\epsilon}\, (\bar{K}^{*0})\right] \vec{\epsilon}\, (\bar{K}^{*0}) \cdot (\vec{p}_{\pi^{+}}-\vec{p}_{K^-}) \cdot G_{\bar{K}^* K}\left( M_{\rm inv} (K^0 \pi^+ K^-)\right) \nonumber\\[1.5mm]
	&&\times T_{K^*\bar{K},K^*\bar{K}}^{(f_1)} \left(M_{\rm inv} (K^0 \pi^+ K^-)\right) \cdot D_{\bar{K}^{*0}}\left(M_{\rm inv}(K^- \pi^+)\right),
\end{eqnarray}
%%%%%%%%%%%%%%%%%
with
%%%%%%%%%%%%%%%%%
\begin{eqnarray}
	D_{K^{*+}}\left(M_{\rm inv}(K^0 \pi^+)\right) &=& \frac{1}{M_{\rm inv}^2(K^0 \pi^+)-m_{K^*}^2+im_{K^*}\Gamma_{K^*}}, \\[2mm]
	D_{\bar{K}^{*0}}\left(M_{\rm inv}(K^- \pi^+)\right) &=& \frac{1}{M_{\rm inv}^2(K^- \pi^+)-m_{\bar{K}^*}^2+im_{\bar{K}^*}\Gamma_{\bar{K}^*}}, \\[2mm]
	\label{eq:tk}
    T_{K^*\bar{K},K^*\bar{K}}^{(f_1)} &=& \frac{g^2_{f_1,K^*\bar{K}}}{M^2_{\rm{inv}}-M_{f_1}^2+iM_{f_1}\Gamma_{f_1}},
\end{eqnarray}
%%%%%%%%%%%%%%%%%
where $\mathcal{C}$ is a constant that contains bare coupling constant of $J/\psi K^* \bar{K}$ or $J/\psi \bar{K}^* K$ vertex and the factor from the $K^{*+} \to K^0 \pi^+$ or $\bar{K}^{*0} \to K^- \pi^+$ coupling.
Summing over the $K^*$ polarizations in Eqs.~\eqref{eq:amp1} and \eqref{eq:amp2}, we have
%%%%%%%%%%%%%%%%%
\begin{eqnarray}
	\label{eq:ta}
	t_a &=& \mathcal{C}\, \vec{\epsilon}\,(J/\psi) \cdot \left[\vec{\epsilon}\,(\omega) \times (\vec{p}_{\pi^+} - \vec{p}_{K^0})\right] \cdot D_{K^{*+}}\left(M_{\rm inv}(K^0 \pi^+)\right) \nonumber\\[2mm]
	&& \times \left[ 1+G_{K^*\bar{K}}\left( M_{\rm inv} (K^0 \pi^+ K^-)\right) \cdot T_{K^*\bar{K},K^*\bar{K}}^{(f_1)} \left(M_{\rm inv} (K^0 \pi^+ K^-)\right)\right], \\[2mm]
	\label{eq:tb}
	t_b &=& \mathcal{C}\, \vec{\epsilon}\,(J/\psi) \cdot \left[\vec{\epsilon}\,(\omega) \times (\vec{p}_{\pi^+} - \vec{p}_{K^-})\right] \cdot D_{\bar{K}^{*0}}\left(M_{\rm inv}(K^- \pi^+)\right) \nonumber\\[2mm]
	&& \times \left[ 1+G_{K^*\bar{K}}\left( M_{\rm inv} (K^0 \pi^+ K^-)\right) \cdot T_{K^*\bar{K},K^*\bar{K}}^{(f_1)} \left(M_{\rm inv} (K^0 \pi^+ K^-)\right)\right].
\end{eqnarray}
%%%%%%%%%%%%%%%%%

We obtain the total amplitude of $J/\psi \to \omega K^0 \pi^+ K^-$ decay by summing $t_a$ and $t_b$ from Eqs.~\eqref{eq:ta} and \eqref{eq:tb}, $\it{i.e}$, $t=t_a+t_b$.
And then the $|t^2|$ is given by
%%%%%%%%%%%%%%%%%
\begin{equation}
	\label{eq:t1}
	|t|^2=|t_a+t_b|^2=|t_a|^2 + |t_b|^2 + 2\,\mathrm{Re}\,(t_a\, t_b^*),
\end{equation}
%%%%%%%%%%%%%%%%%
where $\mathrm{Re}\,(t_a \,t_b^*)$ is the interference term of the two mechanisms in Figs.~\ref{fig-2} and \ref{fig-3}.
In principle, we need to consider this interference term, yet we kown that the $K^0$ and $K^-$ are both in $p$\,-wave in different angles.
Thus, the interference term is neglected between the mechanisms of Figs.~\ref{fig-2} and \ref{fig-3} when we integrate over angles in the differential decay width.
Then Eq.~\eqref{eq:t1} is taken as
%%%%%%%%%%%%%%%%%
\begin{equation}
	\label{eq:t2}
	|t|^2 \simeq |t_a|^2 + |t_b|^2.
\end{equation}
%%%%%%%%%%%%%%%%%

With the decay amplitudes obtained above, we can easily get the decay width of $J/\psi \to \omega K^0 \pi^+ K^-$, which is
%%%%%%%%%%%%%%%%%
\begin{eqnarray}
	\label{eq:Gamfour}
	\Gamma_{J/\psi \to \omega K^0 \pi^{+} K^{-}} &=& \frac{1}{2 M_{J / \psi}} \int \frac{\mathrm{d}^3 p_\omega}{(2 \pi)^3} \cdot \frac{1}{2 E_\omega} \int \frac{\mathrm{d}^3 p_1}{(2 \pi)^3} \;\frac{1}{2 E_1} \int \frac{\mathrm{d}^3 p_2}{(2 \pi)^3} \;\frac{1}{2 E_2} \nonumber\\[2mm]
	&& \times \int \frac{\mathrm{d}^3 p_3}{(2 \pi)^3} \;\frac{1}{2 E_3} \overline{\sum} \sum\;|t|^2\;(2 \pi)^4 \delta\left(p_{J / \psi}^0-E_\omega-E_1-E_2-E_3\right) \nonumber\\[2mm]
	&& \cdot \, \delta^3\left(\vec{p}_{J/\psi}-\vec{p}_\omega-\vec{p}_1-\vec{p}_2-\vec{p}_3\right) \nonumber\\[2mm]
	&=& \frac{1}{2 M_{J / \psi}} \int \frac{\mathrm{d}^3 p_\omega}{(2 \pi)^3} \frac{1}{2 E_\omega} \cdot 2 M_{\text {inv }}\bigg[\frac{1}{2 M_{\text {inv }}} \int \frac{\mathrm{d}^3 p_1}{(2 \pi)^3} \frac{1}{2 E_1} \int \frac{\mathrm{d}^3 p_2}{(2 \pi)^3} \frac{1}{2 E_2} \nonumber\\[2mm]
	&& \times \int \frac{\mathrm{d}^3 p_3}{(2 \pi)^3} \frac{1}{2 E_3} \overline{\sum} \sum|t|^2(2 \pi)^4 \delta\left(p_{J / \psi}^0-E_\omega-E_1-E_2-E_3\right) \nonumber\\[2mm]
	&& \cdot \, \delta^3\left(\vec{p}_{J / \psi}-\vec{p}_\omega-\vec{p}_1-\vec{p}_2-\vec{p}_3\right) \bigg],
\end{eqnarray}
%%%%%%%%%%%%%%%%%
where, $K^0$, $\pi^+$, and $K^-$ are labeled as 1, 2, and 3, respectively, $E_i$ stands for the energy of the $i$ meson, $\vec{p}_i$ stands for the momentum of the $i$ meson, $M_{\rm inv}$ stands for $M_{\rm inv}(K^0\pi^+K^-)$.
The contents of $\left[~~\right]$ in Eq.~\eqref{eq:Gamfour} could be regarded as the three-body decay of a particle with mass equal to $M_{\rm inv}$.
Then Eq.~\eqref{eq:Gamfour}, using the master formula for three body decay in the PDG \cite{Workman:2022ynf},  is represented as
%%%%%%%%%%%%%%%%%
\begin{eqnarray}
	\label{eq:Gam}
	\Gamma_{J / \psi \rightarrow \omega K^0 \pi^{+} K^{-}} &=& \frac{1}{2 M_{J / \psi}} \int \frac{\mathrm{d}^3 p_\omega}{(2 \pi)^3} \;\frac{1}{2 E_\omega} 2 M_{\mathrm{inv}} \int \mathrm{d} M_{12}\, 2 M_{12} \nonumber\\[2mm]
	&& \times \frac{1}{(2 \pi)^3} \cdot \frac{1}{32 M_{\mathrm{inv}}^3} \int \mathrm{d} M_{23}\, 2 M_{23} \overline{\sum} \sum|t|^2,
\end{eqnarray}
%%%%%%%%%%%%%%%%%
where $M_{12}$ and $M_{23}$ stand for $M_{\rm inv}(K^0 \pi^+)$ and $M_{\rm inv}(\pi^+ K^-)$, respectively.
In addition, using $M^2_{\rm inv} = (P_{J/\psi}-p_\omega)^2$, we can write
\begin{eqnarray}
	\label{eq:intdp}
	\int \frac{\mathrm{d}^3 p_\omega}{(2\pi)^3} \;\frac{1}{2E_\omega} &=& \frac{1}{(2\pi)^2}\, p_\omega \,\frac{M_{\rm inv}}{M_{J/\psi}}\, \mathrm{d} M_{\rm inv}, \\[2mm]
	p_\omega &=& \frac{\lambda^{1/2}(M_{J/\psi}^2,m_\omega^2,M_{\mathrm{inv}}^2)}{2M_{J/\psi}}.
\end{eqnarray}
%%%%%%%%%%%%%%%%%

Using Eqs.~\eqref{eq:Gam} and \eqref{eq:intdp}, the differential mass distribution of $J/\psi \to \omega K^0 \pi^+ K^-$ decay is given by
%%%%%%%%%%%%%%%%%
\begin{equation}
    \label{eq:dgamma}
	\begin{aligned}
	\frac{\mathrm{d}\Gamma}{\mathrm{d}M_{\mathrm{inv}}(K^0\pi^+K^-)}=&\frac{1}{(2\pi)^5}\frac{1}{8M_{J/\psi}^2}\frac{1}{M_{\mathrm{inv}}(K^-K^0\pi^+)} p_{\omega}
	\times\int M_{12} \mathrm{d} M_{12}\\
	&\times\int M_{23} \mathrm{d} M_{23}\, \overline{\sum}{\sum}|t^2|,
	\end{aligned}
\end{equation}
%%%%%%%%%%%%%%%%%
where $\overline{\sum}\sum|t^2|=\overline{\sum}\sum|t^2_a| + \overline{\sum}\sum|t^2_b|$.
And $\overline{\sum}\sum|t^2|$ is given by
%%%%%%%%%%%%%%%%%
\begin{eqnarray}
	\label{eq:sumt}
	\overline{\sum}\sum|t^2| &=& \mathcal{D}^2\left|1+G_{K^*\bar{K}}(M_{\rm{inv}}(\pi^+K^-K^0) T^{(f_1)}_{K^*\bar{K},K^*\bar{K}}(M_{\rm{inv}}(\pi^+K^-K^0)\right|^2 \nonumber\\[2mm]
	&& \times\Biggl\{\frac{\lambda(M^2_{\rm{inv}}(\pi^+K^0),m_\pi^2,m_K^2)}{M^2_{\rm{inv}}(\pi^+K^0)}\times\left|\frac{1}{M^2_{\rm{inv}}(\pi^+K^0)-M_{K^*}^2+iM_{K^*}\Gamma_{K^*}}\right|^2 \nonumber\\[2mm]
	&& +\frac{\lambda(M^2_{\rm{inv}}(\pi^+K^-),m_\pi^2,m_K^2)}{M^2_{\rm{inv}}(\pi^+K^-)}\times\left|\frac{1}{M^2_{\rm{inv}}(\pi^+K^-)-M_{K^*}^2+iM_{K^*}\Gamma_{K^*}}\right|^2\Biggr\},
\end{eqnarray}
%%%%%%%%%%%%%%%%%
with $\mathcal{D}^2$ being a constant containing $\mathcal{C}^2$ and factors from other global constants. We take $\mathcal{D}^2 =1$ in our calculations, since we are interested in the line shape of the $K^0\pi^+K^-$ invariant mass distribution.

We need to integrate the invariant masses $M_{12}$ and $M_{23}$ of Eq.~\eqref{eq:dgamma}. The range of $M_{12}$ is given by
%%%%%%%%%%%%%%%%%
\begin{eqnarray}
	M_{12}^{\rm{max}} &=& M_{\rm{inv}}(K^0\pi^+K^-)-m_{K^-},\\[2mm]
    M_{12}^{\rm{min}} &=& m_{K^0}+m_{\pi^+}.
\end{eqnarray}
For fixed $M_{12}$, the limits of $M_{23}$ are given in the PDG as
\begin{eqnarray}	
	M_{23}^{\rm{max}} &=& \sqrt{(E_2^*+E_3^*)^2-\left(\sqrt{E_2^{*2}-m_2^2}-\sqrt{E_3^{*2}-m_3^2}\right)^2},\\[2mm]
    M_{23}^{\rm{min}} &=& \sqrt{(E_2^*+E_3^*)^2-\left(\sqrt{E_2^{*2}-m_2^2}+\sqrt{E_3^{*2}-m_3^2}\right)^2},
\end{eqnarray}
%%%%%%%%%%%%%%%%%
where $E_2^*$ and $E_3^*$ are the energies of $\pi^+$ and $K^-$ in the $M_{12}$ rest frame, respectively,
%%%%%%%%%%%%%%%%%
\begin{eqnarray}
	E_2^* &=& (M_{12}^2-m_1^2+m_2^2)/2M_{12},\\[2mm]
	E_3^* &=& (M^2_{\rm{inv}}(K^-K^0\pi^+)-M_{12}^2-m_3^2)/2M_{12}.
\end{eqnarray}
%%%%%%%%%%%%%%%%%

\subsection{The $J/\psi \to \phi K^0 \pi^+ K^-$ decay}
\label{subsec:phi}

In the same footing as done in the former subsections, we can look at the $\braket{VVP}$ structure and obtain from there the terms containing a $\phi$ and $K^* \bar K$ or $\bar K^* K$. We find now the combination
%%%%%%%%%%%%%%%%%
\begin{equation}\label{eq:Hp1}
	H'=K^{*+} \phi K^-+K^{*0} \phi \bar{K}^0+ \phi K^{*-}K^+ + \phi \bar{K}^{*0}K^0,
\end{equation}
%%%%%%%%%%%%%%%%%
which using the property of the operators of Eq.~\eqref{eq:VVP1} can be cast into
%%%%%%%%%%%%%%%%%
\begin{equation}\label{eq:Hp2}
	H'=-\phi (K^{*+} K^-+K^{*0} \bar{K}^0 - K^{*-}K^+ - \bar{K}^{*0}K^0),
\end{equation}
%%%%%%%%%%%%%%%%%
which is identical to the one of Eq.~\eqref{eq:H} multiplied by $-\sqrt{2}$ (see also Ref.~\cite{Jiang:2019ijx}).

The differential width for $\phi$ production, instead of $\omega$, is easily obtained with the same formulas, simply changing the mass of the $\omega$ to the one of the $\phi$ and multiplying by $2$ the differential width of Eq.~\eqref{eq:dgamma}.

%=====================
%=====================
%=====================
\section{Results}
\label{sec:result}
%=====================
%=====================
%=====================

\begin{figure}[t]
	\includegraphics[width=0.7\linewidth]{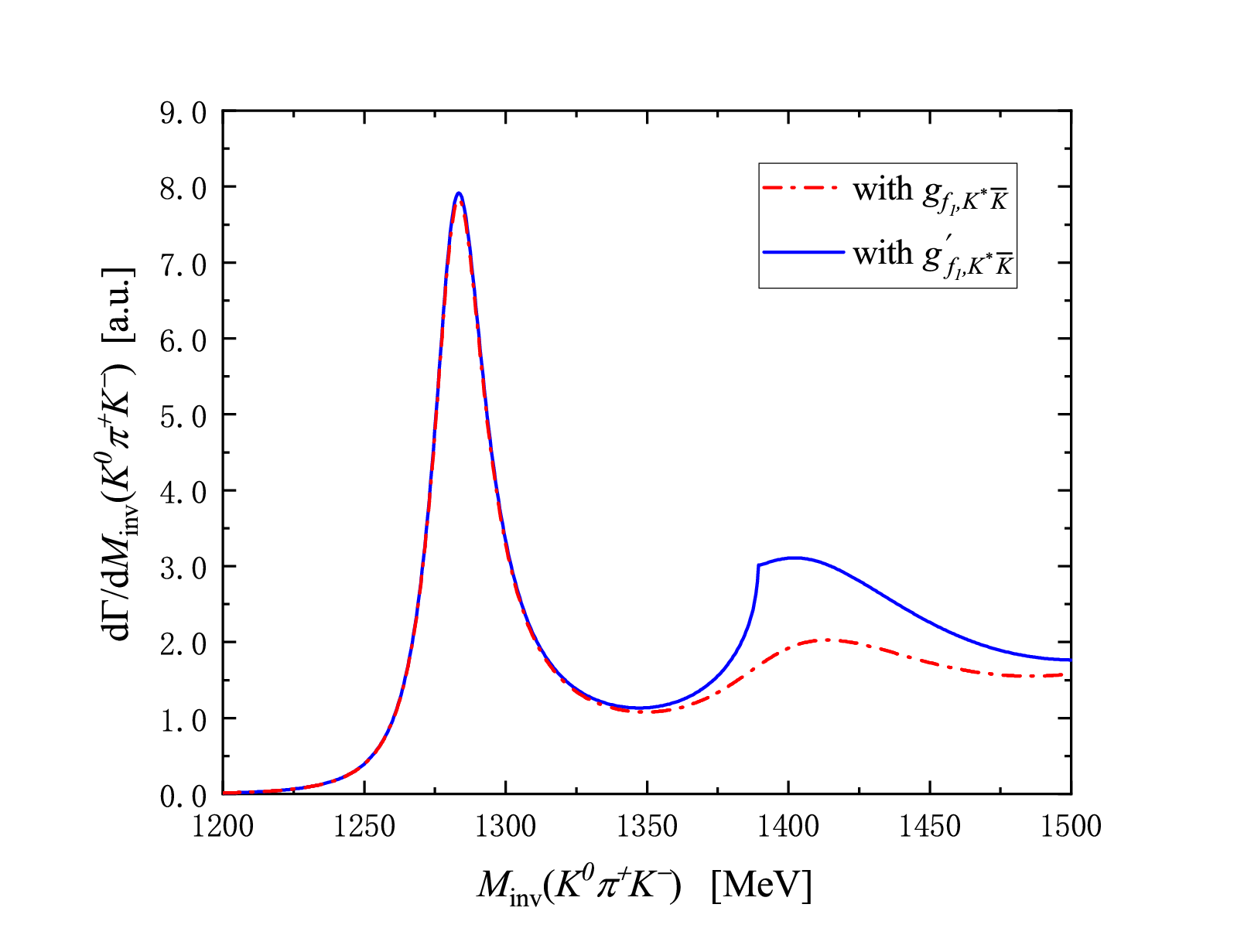}
	\vspace{-0.4cm}
	 \caption{$M_{\mathrm{inv}}(K^0\pi^+ K^-)$ mass distribution for $J/\psi \to \omega K^0 \pi^+ K^-$ decay with the couplings $g_{f_1,K^*\bar{K}} = 7474-i 364 \mev$ and $g_{f_1,K^*\bar{K}}^\prime = 7469 \mev$.}
	 \label{fig:result}
\end{figure}

The $K^0 \pi^+ K^-$ invariant mass distribution  $\frac{\mathrm{d}\Gamma}{\mathrm{d}M_{\mathrm{inv}}(K^0\pi^+ K^-)}$ of the $J/\psi \to \omega K^0 \pi^+ K^-$ decay is shown in Fig.~\ref{fig:result},
%%%%%%%%%%%%%%%%%
 %%%%%%%%%%%%%%%%%
 where the blue line corresponds to the case considering the width of $K^*$ and having the coupling $g_{f_1,K^*\bar{K}} = 7474 -i 364 \mev$, while the red dash-dot line corresponds to the case without considering the width of $K^*$ and having the coupling $g_{f_1,K^*\bar{K}}^\prime = 7469 \mev$.

It is interesting to compare the shapes of these two cases in Fig.~\ref{fig:result}.
We observe two peaks in both cases, but a little different for the shape of the second peak.
There is a clear signal for the $f_1(1285)$ production, and the curves of these two cases are almost the same.
Their mass and width are about $1283 \mev$ and $23 \mev$, respectively, which are consistent with the $f_1(1285)$ of PDG \cite{Workman:2022ynf}.
The second peak is observed around $1420 \mev$.
We can make a rough estimate of the width for the second peak, which could be done by  drawing a straight line between the minimum at $1350 \mev$ and the last point of the distribution.
We obtain the width of $70 \mev$, a bit larger than the $55 \mev$ given by PDG \cite{Workman:2022ynf}.
The width of this peak qualitatively agrees with the average value given by PDG considering the approximations done in the model, and the dispersion of values between different experiments tabulated in the PDG \cite{Workman:2022ynf}.
However, neither parameters nor information about the $f_1(1420)$ are introduced in the formalism of Sec.~\ref{sec:form}.
We can also see that the second peak has a cusp in the case without taking into account the width of $K^*$ meson, corresponding to the opening of the $K^* \bar K$ or $\bar K^* K$ channels, while the curve becomes flatter with the width of $K^*$ meson considered.

It is interesting to see the origin of this second peak. From the molecular perspective of the $f_1(1285)$ as a $K^* \bar K +c.c.$ bound state, the $f_1(1285)$ cannot decay to $K^* \bar K$. Yet, since the resonance has a width, for higher values of its mass distribution the resonance has enough energy to decay to the $K^* \bar K$ or $\bar K^* K$.
This is also favored by the $K^*$ width, which leads to a $K^*$ mass distribution, with mass components below the nominal mass.
As a consequence, we find the combination of two factors: the opening of the $K^* \bar K$ decay channel and the reduced strength of the $f_1(1285)$ mass distribution at its tail.
The combination of the two factors leads to the peak that we have calculated.
What we find is that the peak observed around $1420 \,\mev$ in this and other experiments \cite{Debastiani:2016xgg} is not due to a genuine resonance, but an unavoidable consequence of the decay of the $f_1(1285)$ into $\bar K^* K, K^* \bar K$, once the $f_1(1285)$ is considered as a molecular state of these components.
However, this latter condition is not necessary. It is enough that the $f_1(1285)$ resonance has a reasonable coupling to $K^* \bar K$ for this peak to appear.
Then this peak would correspond to the $f_1(1285)$ decay to $\bar K^* K, K^* \bar K$, which appears at a different energy than the nominal mass of the $f_1(1285)$ for the reasons that we have discussed above.

\begin{figure}[t]
	\includegraphics[width=0.65\linewidth]{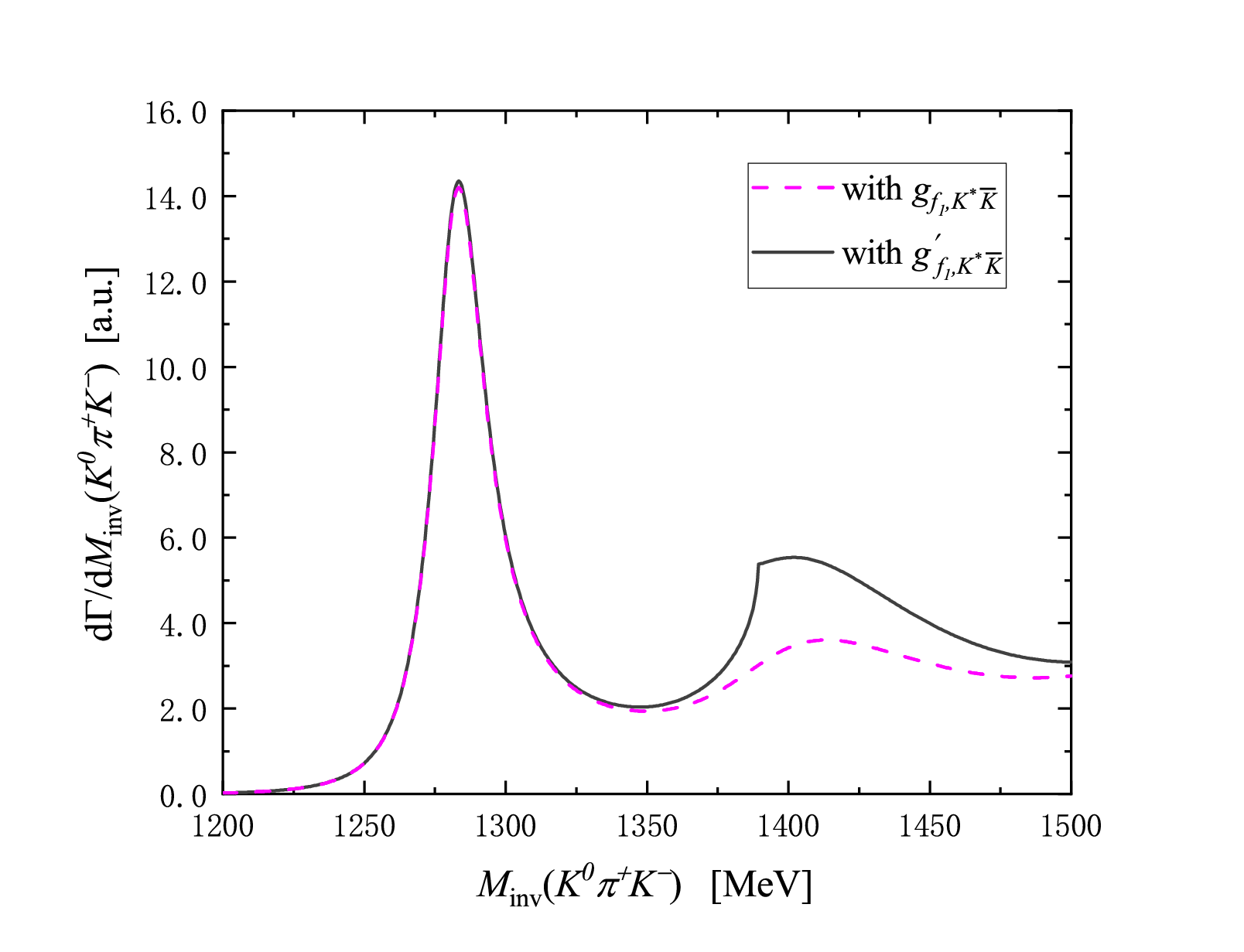}
	\vspace{-0.4cm}
	 \caption{$M_{\mathrm{inv}}(K^0\pi^+ K^-)$ mass distribution for $J/\psi \to \phi K^0 \pi^+ K^-$ decay with the couplings $g_{f_1,K^*\bar{K}} = 7474-i 364 \mev$ and $g_{f_1,K^*\bar{K}}^\prime = 7469 \mev$.}
	 \label{fig:f1phi}
\end{figure}

\begin{figure}[t]
	\includegraphics[width=0.65\linewidth]{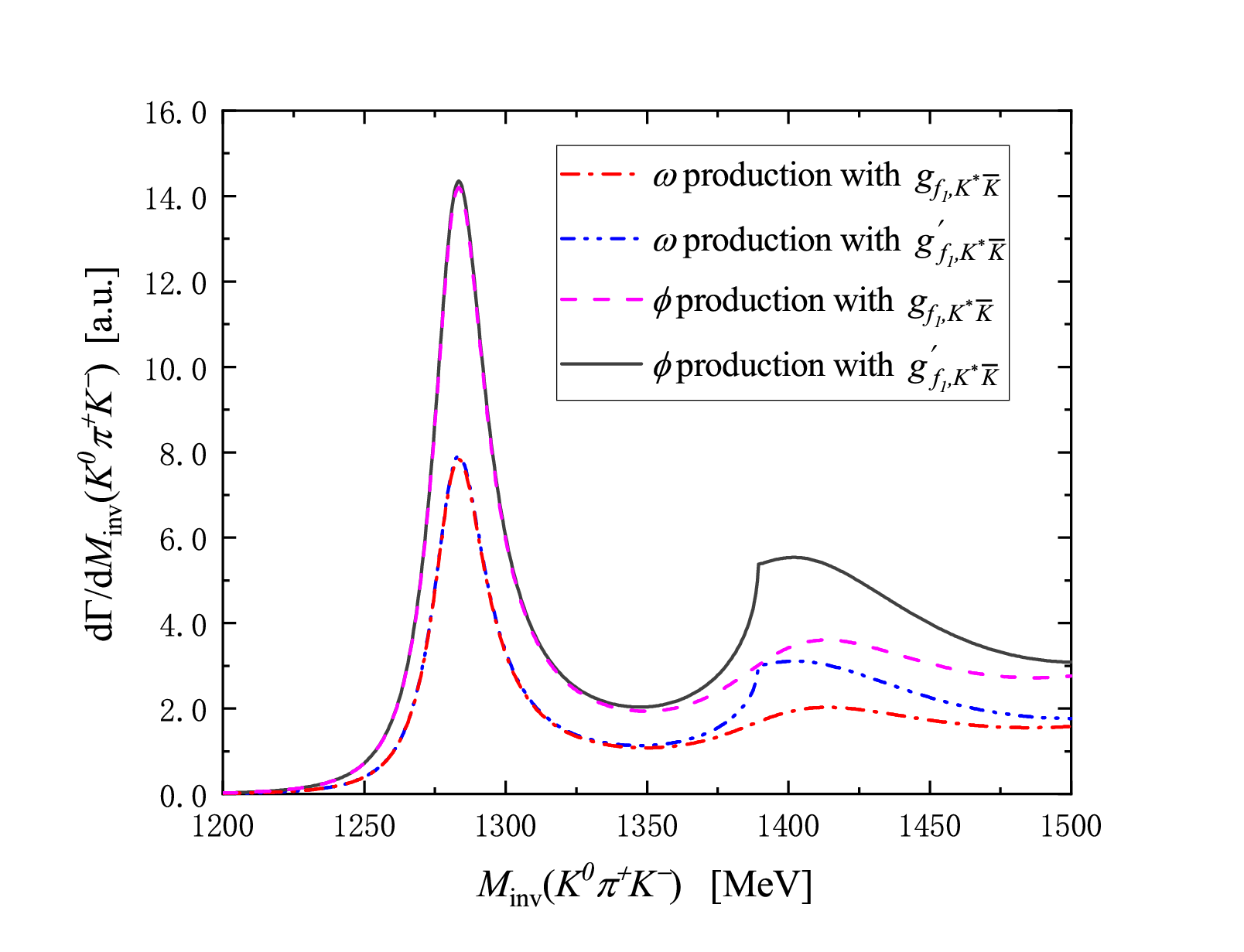}
	\vspace{-0.4cm}
	 \caption{Comparison of the $M_{\mathrm{inv}}(K^0\pi^+ K^-)$ mass distribution for $J/\psi \to \omega K^0 \pi^+ K^-$ and $J/\psi \to \phi K^0 \pi^+ K^-$ decays.}
	 \label{fig:f1total}
\end{figure}

We show the results for the $K^0 \pi^+ K^-$ invariant mass distribution $\frac{\mathrm{d}\Gamma}{\mathrm{d}M_{\mathrm{inv}}(K^0\pi^+ K^-)}$ of the $J/\psi \to \phi K^0 \pi^+ K^-$ decay in Fig.~\ref{fig:f1phi}, and in Fig.~\ref{fig:f1total} we show results for $\omega$ and $\phi$ production overlayed.
%%%%%%%%%%%%%%%%%
 %%%%%%%%%%%%%%%%%
We see that the results are very similar to the ones for the $J/\psi \to \omega K^0 \pi^+ K^-$ decay. We note that there are two different factors: the phase space that favors $\omega$ production and the factor $(-\sqrt{2})$ in the amplitude.
As a consequence, we obtain a ratio for the production of $\phi$ to $\omega$ in this region of about $1.8$, a little smaller than $(\sqrt{2})^2$.

We should comment that in Ref.~\cite{MARK-III:1986unh} these two decay modes were measured, albeit with poor statistics, concluding that an enhancement around $1.44\, \gev$ is observed in the $K^+ K^- \pi^0$ invariant mass distribution with $\omega$ production, but no signal is seen at the nominal mass of the $f_1(1285)$. The reverse pattern seems to appear for $\phi$ production, which shows an enhancement around $1.28 \, \gev$ and no evidence for structure at $1.4\, \gev$.
We find from our calculation that both signals appear in $\phi$ as well as $\omega$ production. In view of these contradicting findings and the poor statistics in the experiment of Ref.~\cite{MARK-III:1986unh}, it would be most opportune to have a future look at these decays with much improved statistics.

%=====================
%=====================
%=====================
\section{Conclusions}
\label{sec:concl}
%=====================
%=====================
%=====================

In the present work, based on the perspective that the $f_1(1285)$ is dynamically generated from a single channel $K^* \bar{K} +c.c.$ interaction, we have carried out of a study of the $J/\psi \to \omega (\phi) K^{*+} K +c.c. \to \omega (\phi) K^0 \pi^+ K^-$ decay.
Both the tree level diagrams and the rescattering diagrams of $K^* \bar{K}$ and $\bar{K}^* K$ components to produce the $f_1(1285)$ are taken into account.
The results, with and without the width of the $K^*$ meson in the loop function, are shown when calculating the differential mass distribution of the $J/\psi \to \omega (\phi) K^0 \pi^+ K^-$.

We show the results for the $K^0 \pi^+ K^-$ invariant mass distribution of the $J/\psi \to \omega K^0 \pi^+ K^-$ decay in Fig.~\ref{fig:result} and in Fig~\ref{fig:f1phi} for $J/\psi \to \phi K^0 \pi^+ K^-$ decay.
Two peaks in the $K^0 \pi^+ K^-$ mass distribution are observed, one clear peak at the $f_1(1285)$ nominal mass corresponds to the production of the $f_1(1285)$ in the $J/\psi \to \omega K^0 \pi^+ K^-$ process, and the other peak locates at around $1420 \mev$ with a width of about $70 \mev$. The latter peak is consistent with the $f_1(1420)$ quoted in PDG \cite{Workman:2022ynf}.
However, neither parameters nor information about the $f_1(1420)$ are introduced in our theoretical calculations.
In the framework of our study, the $f_1(1420)$ observed in experiment is not a genuine resonance, this peak is a manifestation of the kinematic effect in a higher energy region caused by the $K^* \bar{K} + c.c.$ decay mode of the $f_1(1285)$.
The $J/\psi \to \omega f_1(1420)$ reaction provides an important way to study the nature of the $f_1(1420)$, hence more accurate experimental data are needed to verify the research results in this paper.

Altogether, our study indicates that the observation of the $f_1(1285)$ in the decays of $J/\psi \to \omega K^0 \pi^+ K^-$ and $J/\psi \to \phi K^0 \pi^+ K^-$ provides relevant information concerning the nature of the axial vector mesons.
The observation of ``$f_1(1420)$'' peak gives a motivation and a new perspective for future experimental measurements.

%=====================
%=====================
%=====================
\begin{acknowledgments}
	This work is partly supported by the National Natural Science Foundation of China (NSFC) under Grants No. 11975083, No. 12365019 and No. 12075019, and by the Central Government Guidance Funds for Local Scientific and Technological Development, China (No. Guike ZY22096024), the Jiangsu Provincial Double-Innovation Program under Grant No.~JSSCRC2021488, and the Fundamental Research Funds for the Central Universities.
	This work is also partly supported by the Spanish Ministerio de Economia y Competitividad (MINECO) and European FEDER
	funds under Contracts No. FIS2017-84038-C2-1-P B, PID2020-112777GB-I00, and by Generalitat Valenciana under contract
	PROMETEO/2020/023.
	This project has received funding from the European Union Horizon 2020 research and innovation
	programme under the program H2020-INFRAIA-2018-1, grant agreement No. 824093 of the STRONG-2020 project.
	
\end{acknowledgments}

%\clearpage

\end{document}